\pdfoutput=1

\documentclass[11pt,twoside,a4paper,cmspaper,final,collab]{cms-tdr}
\def\svnVersion{267149}\def\cmsCernNo{2013-037}\def\cmsCernDate{\today}\def\cmsMessage{Published in Physics Letters B as \href{http://dx.doi.org/10.1016/j.physletb.2014.10.063}{\doi{10.1016/j.physletb.2014.10.063.}}}

\begin{document}\cmsNoteHeader{EXO-12-032}

\hyphenation{had-ron-i-za-tion}
\hyphenation{cal-or-i-me-ter}
\hyphenation{de-vices}
\RCS$Revision: 266510 $
\RCS$HeadURL: svn+ssh://svn.cern.ch/reps/tdr2/papers/EXO-12-032/trunk/EXO-12-032.tex $
\RCS$Id: EXO-12-032.tex 266510 2014-11-04 21:16:03Z pedrok $
\newlength\cmsFigWidth
\ifthenelse{\boolean{cms@external}}{\setlength\cmsFigWidth{0.95\columnwidth}}{\setlength\cmsFigWidth{0.65\textwidth}}
\ifthenelse{\boolean{cms@external}}{\providecommand{\cmsLeft}{top}}{\providecommand{\cmsLeft}{left}}
\ifthenelse{\boolean{cms@external}}{\providecommand{\cmsRight}{bottom}}{\providecommand{\cmsRight}{right}}
\newcommand{\ST}{\ensuremath{S_\mathrm{T}}\xspace}
\newcommand{\tauh}{\ensuremath{\tau_\mathrm{h}}\xspace}
\newcommand{\etau}{\ensuremath{\Pe\tauh}\xspace}
\newcommand{\mutau}{\ensuremath{\Pgm\tauh}\xspace}
\newcommand{\elltau}{\ensuremath{\ell\tauh}\xspace}
\newcommand{\elltaubj}{\ensuremath{\ell\tauh \cPqb j}\xspace}
\newcommand{\emu}{\ensuremath{\Pe\Pgm}\xspace}
\newcommand{\MassTJ}{\ensuremath{M(\tauh,\text{jet})}\xspace}
\newcommand{\MassTJone}{\ensuremath{M(\tauh,j_{1})}\xspace}
\newcommand{\MassTJtwo}{\ensuremath{M(\tauh,j_{2})}\xspace}
\newcommand{\MassTJm}{\ensuremath{M(\tauh,j_{m})}\xspace}
\newcommand{\MassLJone}{\ensuremath{M(\ell,j_{1})}\xspace}
\newcommand{\MassLJtwo}{\ensuremath{M(\ell,j_{2})}\xspace}
\newcommand{\MassLJn}{\ensuremath{M(\ell,j_{n})}\xspace}
\newcommand{\Zmm}{\ensuremath{\cPZ \to \Pgm\Pgm}\xspace}%
\newcommand{\sTauR}{\ensuremath{\widetilde{\tau}_\cmsSymbolFace{R}}\xspace}

\cmsNoteHeader{EXO-12-032}
\title{Search for pair production of third-generation scalar leptoquarks and top squarks in proton-proton collisions at $\sqrt{s} = 8\TeV$}

\date{\today}

\abstract{
A search for pair production of third-generation scalar leptoquarks
and supersymmetric top quark partners, top squarks, in final states involving
tau leptons and bottom quarks is presented.
The search uses events from a data sample of proton-proton collisions
corresponding to an integrated luminosity of 19.7\fbinv, collected
with the CMS detector at the LHC with $\sqrt{s}=8$\TeV.
The number of observed events
is found to be in agreement with the expected standard model background.
Third-generation scalar leptoquarks with masses
below 740\GeV are excluded at 95\% confidence level,
assuming a 100\% branching fraction
for the leptoquark decay to a tau lepton and a bottom quark.
In addition, this mass limit applies directly to
top squarks decaying via an R-parity violating coupling $\lambda^{\prime}_{333}$.
The search also considers a similar signature from top squarks
undergoing a chargino-mediated decay involving the R-parity violating coupling $\lambda^{\prime}_{3jk}$.
Each top squark decays to a tau lepton, a bottom quark, and two light quarks.
Top squarks in this model with masses below 580\GeV are excluded at
95\% confidence level. The constraint on the leptoquark mass is the most stringent to date,
and this is the first search for top squarks decaying via $\lambda^{\prime}_{3jk}$.
}

\hypersetup{%
pdfauthor={CMS Collaboration},%
pdftitle={Search for pair production of third-generation scalar leptoquarks and top squarks in proton-proton collisions at sqrt(s) = 8 TeV},%
pdfsubject={CMS},%
pdfkeywords={CMS, physics, leptoquark, top squark}}

\maketitle

\section{Introduction}
Many extensions of the standard model (SM)
predict new scalar or vector bosons, called leptoquarks (LQ),
which carry non-zero lepton and baryon numbers, as well
as color and fractional electric charge.
Examples of such SM extensions include
SU(5) grand unification~\cite{GUT},
Pati--Salam SU(4)~\cite{SU4}, composite models~\cite{LQ3b},
superstrings~\cite{SUPERSTR}, and technicolor models~\cite{TC3}.
Leptoquarks decay into a quark and a lepton, 
with a model-dependent branching fraction for each possible decay.
Experimental limits on flavor-changing neutral
currents and other rare processes
suggest that searches should focus on
leptoquarks that couple to quarks and leptons within
the same SM generation, for leptoquark masses accessible to current
colliders~\cite{LQ1,LQ3b}.

The dominant pair production mechanisms for leptoquarks
at the CERN LHC would be gluon-gluon fusion
and quark-antiquark annihilation via quantum chromodynamic (QCD)
couplings.
The cross sections for these processes
depend only on the leptoquark mass for scalar leptoquarks.
In this Letter, a search with the CMS detector
for third-generation scalar leptoquarks,
each decaying to a tau lepton and a bottom quark, is presented.

Similar signatures arising from supersymmetric models
are also covered by this search.
Supersymmetry (SUSY)~\cite{SUSY1,SUSY2}
is an attractive extension of the SM because it can resolve the
hierarchy problem without unnatural fine-tuning,
if the masses of the supersymmetric partner of the top quark (top squark)
and the supersymmetric partners of the Higgs boson (higgsinos) are not too
large~\cite{Papucci:2011wy,SUSYNATURAL}.
In many natural SUSY models
the top squark and the higgsinos are substantially lighter
than the other scalar SUSY particles.
This light top squark scenario can be realized
in both R-parity conserving (RPC)
and R-parity violating (RPV) SUSY models,
where R-parity is a new
quantum number~\cite{Barbier20051}
that distinguishes SM and SUSY particles.
In the context of an RPC decay of the top squark,
the presence of an undetected particle
(the lightest SUSY particle) is expected to generate a
signature with large missing transverse momentum.
If R-parity is violated, however, SUSY particles can decay
into final states containing only SM particles.
The RPV terms in the superpotential are:
\begin{equation}
W \ni \frac{1}{2}\lambda_{ijk}L_{i}L_{j}E_{k}^{c} + \lambda_{ijk}^{\prime}L_{i}Q_{j}D_{k}^{c} + \frac{1}{2}\lambda_{ijk}^{\prime \prime}U_{i}^{c}D_{j}^{c}D_{k}^{c} + \mu_{i}L_{i}H_{u}
\end{equation}
where $W$ is the superpotential; $L$ is the lepton doublet superfield; $E$ is the lepton singlet superfield; $Q$ is the quark doublet superfield; $U$ and $D$ are the quark singlet superfields; $H_{u}$ is the Higgs doublet superfield that couples to up-type quarks; $\lambda$, $\lambda^{\prime}$, and $\lambda^{\prime \prime}$ are coupling constants; and $i$, $j$, and $k$ are generation indices.

At the LHC, top squarks (${\sTop}$) would be directly pair-produced
via strong interactions. In this search, two different
decay channels of directly produced top squarks are considered.
Both scenarios relate to simplified models
in which all of the other SUSY particles have masses
too large to participate in the interactions.
In the first case we study the two-body lepton number violating decay
$\sTop\to\tau \cPqb$~\cite{Barbier20051} with a coupling constant
$\lambda^{\prime}_{333}$ allowed by the trilinear RPV operators.
The final-state signature and kinematic distributions of such a signal are
identical to those from the pair production of third-generation
scalar leptoquarks. When the masses of the supersymmetric
partners of the gluon and quarks, excluding the top squark,
are large, the top squark pair production cross section is
the same as that of the third generation LQ.
Thus, the results of the leptoquark search can be directly
interpreted in the context of RPV top squarks.

In some natural SUSY models~\cite{Jared}, if
the higgsinos (\chiz, \chipm) are lighter than the top squark, or if the RPV
couplings that allow direct decays to SM particles are sufficiently small,
the top squark decay may preferentially proceed via superpartners.
In the second part of the search we focus on a scenario
in which the dominant RPC decay of the top squark is $\sTop\to\chipm\cPqb$.
This requires the mass splitting between the top squark and the chargino
to be less than the mass of the top quark, so it is chosen to be 100\GeV.
The chargino is assumed to be a pure higgsino
and to be nearly degenerate in mass with the neutralino.
We consider the case when
$\chipm\to\sNu\tau^{\pm} \to \cPq\cPq\tau^{\pm}$.
The decay of the sneutrino occurs according to an RPV operator with a coupling
constant $\Lamp_{3jk}$, where the cases $j, k = 1, 2$ are considered.
Such signal models can only be probed by searches that do not require
large missing transverse momentum, as
the other decay of the chargino, $\chipm\to\nu\sTau$, does not contribute
to scenarios involving the $\Lamp_{3jk}$ coupling because of chiral suppression.
From such a signal process, we expect events with
two tau leptons, two jets originating from hadronization of the bottom quarks,
and at least four additional jets.

In this Letter, the search for scalar
leptoquarks and top squarks decaying through the coupling $\lambda^{\prime}_{333}$
is referred to as the leptoquark search.
The search for the chargino-mediated decay of top squarks involving the $\Lamp_{3jk}$
coupling is referred to as the top squark search.
The data sample used in this search has been recorded with the CMS detector
in proton-proton collisions at a center-of-mass energy of 8\TeV
and corresponds to an integrated luminosity of 19.7\fbinv.
One of the tau leptons in the final state is required
to decay leptonically: $\tau\to \ell\bar{\cPgn}_{\ell} \cPgn_{\tau}$,
where $\ell$ can be either an electron or a muon,
denoted as a light lepton.
The other tau lepton is required
to decay to hadrons ($\tauh$):
$\tau \to \text{hadrons} + \nu_{\tau}$.
These decays result in two possible final states labeled below
as $\etau$ and $\mutau$, or collectively $\elltau$ when the lepton flavor is unimportant.
The leptoquark search is performed in a mass range from 200 to 1000\GeV using a sample of
events containing one light lepton, a hadronically decaying tau lepton, and
at least two jets, with at least one of the jets identified as
originating from bottom quark hadronization (b-tagged).
The top squark search is performed in a mass range from 200 to 800\GeV using a
sample of events containing one light lepton, a hadronically decaying tau lepton, and
at least five jets, with at least one of the jets b-tagged.

No evidence for third-generation leptoquarks or
top squarks decaying to tau leptons and bottom quarks
has been found in previous search\-es \cite{CMSLQ3,ATLASLQ3}.
The most stringent lower limit to date on the mass of a scalar third
generation leptoquark decaying to a tau lepton and a bottom quark
with a 100\% branching fraction is about 530\GeV from both the
CMS and ATLAS experiments. This Letter also presents the first search for the chargino-mediated decay
of the top squark through the RPV coupling $\Lamp_{3jk}$.

\section{The CMS detector}

The central feature of the CMS
apparatus is a superconducting solenoid, of 6\unit{m} internal diameter, providing
a field of 3.8\unit{T}.  Within the field volume are several subdetectors.
A silicon pixel and strip tracker allows the reconstruction of the trajectories
of charged particles within the pseudorapidity range $\abs{\eta}<2.5$.
The calorimetry system consists of a lead tungstate crystal electromagnetic
calorimeter (ECAL) and a brass and scintillator hadron calorimeter,
and measures particle energy depositions for $\abs{\eta} < 3$.
The CMS detector also has extensive forward calorimetry ($2.8<\abs{\eta}<5.2$).
Muons are measured in gas-ionization detectors embedded in the steel flux-return yoke of the magnet.
Collision events are selected using a two-tiered trigger system~\cite{CMSTrigger}.
A more detailed description of the CMS detector,
together with a definition of the coordinate system used and the relevant kinematic variables,
can be found in Ref.~\cite{CMSdetector}.

\section{Object and event selection}

Candidate LQ or top squark events were collected using a set of triggers
requiring the presence of either an electron or a muon
with transverse momentum (\pt) above a threshold of 27 or 24\GeV, respectively.

Both electrons and muons are required to be reconstructed within the range $\abs{\eta}<2.1$
and to have $\pt>30\GeV$.
Electrons, reconstructed using information from the ECAL and the tracker, are required to have
an electromagnetic shower shape consistent with
that of an electron, and an energy deposition in ECAL that is compatible
with the track reconstructed in the tracker.
Muons are required to be reconstructed
by both the tracker and the muon spectrometer.
A particle-flow (PF)
technique~\cite{Pflow,CMS-PAS-PFT-10-002,CMS-PAS-PFT-10-003} is
used for the reconstruction of hadronically decaying tau lepton
candidates.
In the PF approach, information from all subdetectors
is combined to reconstruct
and identify all final-state particles produced in the collision.
The particles are classified as either charged hadrons, neutral hadrons,
electrons, muons, or photons.  These particles are used with
the ``hadron plus strips'' algorithm~\cite{HPS} to identify $\tauh$ objects.
Hadronically decaying tau leptons with one or
three charged pions and up to two neutral pions are reconstructed.
The reconstructed $\tauh$
is required to have visible $\pt> 50\GeV$ and
$\abs{\eta}<2.3$.
Electrons, muons, and tau leptons are required to
be isolated from other reconstructed particles.
The identified electron (muon) and $\tauh$ are required to
originate from the same vertex and be separated
by $\Delta R = \sqrt{\smash[b]{(\Delta\eta)^2 + (\Delta\phi)^2}} > 0.5$.
The light lepton and the $\tauh$ are also required to
have opposite electric charge.
Events are vetoed if another light lepton is found, passing
the kinematic, identification, and isolation criteria described above,
that has an opposite electric charge from the selected light lepton.

Jets are reconstructed using the anti-\kt algorithm~\cite{Salam:2009jx,Cacciari:2008gp} with a
size parameter 0.5
using particle candidates reconstructed with the PF technique. Jet energies are corrected
by subtracting the average contribution from particles coming from
other proton-proton collisions in the same beam crossing (pileup)
and by applying a jet energy calibration, determined empirically~\cite{CMS-JEC}.
Jets are required to be within $\abs{\eta}<2.4$,
have $\pt>30\GeV$, and be
separated from both the light lepton and the $\tauh$ by
$\Delta R>0.5$. The minimum jet \pt requirement eliminates most jets from pileup interactions.
Jets are b-tagged using the
combined secondary vertex algorithm with the loose operating point~\cite{BTV-12-001}.
In the leptoquark search, the b-tagged jet with the highest \pt is selected, and then the remaining jet
with the highest \pt is selected whether or not it is b-tagged.
In the top squark search, the b-tagged jet with the highest \pt is selected, and then the remaining four jets
with the highest \pt are selected whether or not they are b-tagged.

To discriminate between signal and background in the leptoquark search,
the mass of the $\tauh$ and a jet, denoted $\MassTJ$,
is required to be greater than 250\GeV.
There are two possible pairings of the $\tauh$ with the two required jets.
The pairing is chosen to minimize the difference between the mass
of the $\tauh$ and one jet
and the mass of the light lepton and another jet. According to a simulation,
the correct pairing is selected in approximately 70\% of events.

The \ST distribution after the final selection is used to extract the limits
on both the leptoquark and top squark signal scenarios, where \ST is defined as the scalar
sum of the \pt of the light lepton, the $\tauh$, and the two jets (five jets) for
the leptoquark search (top squark search).

\section{Background and signal models}

Several SM processes can mimic the final-state signatures expected
from leptoquark or top squark pair production and decay.
For this analysis, the backgrounds are divided into three groups, which are denoted as
\ttbar irreducible, major reducible, and other.
The \ttbar irreducible background comes from the pair production of top quarks (\ttbar)
when both the light lepton and \tauh are genuine, each produced from the decay of a W boson.
In this case, the light lepton can originate either directly from the W boson decay or
from a decay chain $\PW\to\tau  \cPgn_{\tau} \to \ell\cPgn_{\ell} \cPgn_{\tau} \cPgn_{\tau}$.
The major reducible background consists of events in which a quark or gluon jet is
misidentified as a \tauh.
The processes contributing to the major reducible background are
associated production of a W or Z boson with jets, and \ttbar.
Additionally, a small contribution from the QCD multijet process is included, in which
both the light lepton and the \tauh are misidentified jets.
The third group, other backgrounds, consists of processes that make small contributions
and may contain either genuine or misidentified tau leptons.
This includes the diboson and single-top-quark processes,
the \ttbar and \cPZ+jets processes when a light lepton is
misidentified as a \tauh, and the \cPZ+jets
process when the Z boson decays to a pair of tau leptons.
The other backgrounds are estimated from the simulation described below, while the \ttbar irreducible
and major reducible backgrounds are estimated using observed data.
The major reducible and other backgrounds include events with 
both genuine and misidentified light leptons.

The \PYTHIA v6.4.24 generator~\cite{Sjostrand:2006za}
is used to model the signal and diboson processes.
The leptoquark signal samples are generated with masses ranging from 200 to 1000\GeV, and
the top squark signal samples are generated with masses ranging from 200 to 800\GeV
and the sneutrino mass set to 2000\GeV.
The \MADGRAPH v5.1.3.30 generator~\cite{MadGraph} is used to model the
\ttbar, \PW+jets, and \cPZ+jets processes.
This generation includes contributions from heavy-flavor and extra jets.
The single top-quark production is modeled with the \POWHEG 1.0 r138~\cite{POWHEG2,POWHEG:singlet,POWHEG:singletW}
generator.
Both the \MADGRAPH and \POWHEG generators
are interfaced with \PYTHIA for hadronization and showering.
The \TAUOLA program~\cite{TAUOLA} is used for tau lepton decays in
the leptoquark, \ttbar, \PW+jets, \cPZ+jets, diboson, and single top-quark samples.
Each sample is passed through a full simulation of the CMS detector
based on \GEANTfour~\cite{Agostinelli2003250} and the complete set of reconstruction algorithms is used to analyze collision data.
Cross sections for the leptoquark signal and diboson processes
are calculated to next-to-leading order (NLO)~\cite{LQ-sigmaNLO,MCFM}.
The cross sections for the top squark signal are calculated at
NLO in the strong coupling constant, including the resummation of soft gluon emission at
next-to-leading-logarithmic accuracy
(NLO+NLL)~\cite{bib-nlo-nll-01,bib-nlo-nll-02,bib-nlo-nll-03,bib-nlo-nll-04,bib-nlo-nll-05}.
The next-to-next-to-leading-order or
approximate next-to-next-to-leading-order~\cite{FEWZ,TOPCrossSec}
cross sections are used for the rest of the background processes.

The efficiencies of the trigger and final selection criteria
for signal processes are
estimated from the simulation.
The efficiencies for light leptons and
\cPqb~jets are calculated from data
and used where necessary to correct the event selection efficiency
estimations from the simulation.
No correction is required for hadronically decaying tau leptons.

The \ttbar irreducible background is estimated from an $\emu$ sample
that is 87\% pure in \ttbar events according to the simulation.
The contributions from other processes are simulated and subtracted from the observed data.
This sample comprises events with one electron and one muon
that satisfy the remaining final selection criteria, except that a \tauh is not required.
The potential signal contamination of this sample has been found
to be negligible for any signal mass hypothesis.
The final yield of the \emu sample is scaled by the relative difference in the selection efficiencies
between the \elltau and \emu samples. The selection efficiencies are measured in the
simulation and are corrected to match those from collision data.
The estimation of the final yield based on the observed data agrees with both the direct prediction
from the simulation and the yield obtained after applying the same method to the
Monte Carlo (MC) samples.
The \ST distribution for the \ttbar irreducible background is obtained
from a simulated \ttbar sample
that consists exclusively of fully leptonic decays of top quarks.

The major reducible background from \ttbar, \PW+jets, and \cPZ+jets events in
which a jet is misidentified as a hadronically decaying tau lepton
is estimated from observed data. The probability of misidentification is measured
using events recorded with a \cPZ~boson produced in association with jets
and decaying to a pair of muons (\Zmm).
The invariant mass of the muon pair is required to be greater than 50\GeV
and events are required to contain at least one jet that is incorrectly identified as a \tauh
and may or may not pass the isolation requirement.
The misidentification probability $f(\pt(\tau))$ is calculated as the fraction of these \tauh
candidates that pass the isolation requirement
and depends on the \pt of the candidates.
The background yield is estimated from a sample of events
satisfying the final selection criteria, except that all
\tauh candidates in the events must fail the isolation requirement.
Equation \eqref{eq:faketauest} relates the yield of these ``anti-isolated''
events to the yield of events passing the final selection,
using the misidentification probability:
\begin{equation}
  N_{\text{misID $\tau$}} = \sum_{\text{events}}^{\text{(anti-iso)}} \frac{1 - \prod_{\tau}\big[1-f\big(\pt(\tau)\big)\big]}{\prod_{\tau}\big[1-f\big(\pt(\tau)\big)\big]}.
  \label{eq:faketauest}
\end{equation}
The estimation of the final yield based on the observed data agrees with both the direct prediction from the simulation
and the estimation performed using the same approach on simulated samples.
The \ST distribution for the major reducible background is obtained using simulated samples for
the \PW+jets and \cPZ+jets processes and the \ttbar process with exclusively semi-leptonic decays.

The QCD multijet process contributes only in the \etau channel in the leptoquark search and corresponds to 16\% of the reducible background.
The contribution from multijet events is estimated from a sample
of observed events satisfying the final selection criteria for the \etau channel except that
the electron and \tauh must have the same electric charge.
The QCD component is included in the distribution of
the rest of the major reducible background, described above.

\section{Systematic uncertainties}

There are a number of systematic uncertainties associated with
both the background estimation and the signal efficiency.
The uncertainty in the total integrated luminosity is
2.6\%~\cite{CMS-LUMI}.
The uncertainty in the trigger and lepton efficiencies
is 2\%, while the uncertainty assigned to the $\tauh$
identification efficiency is 6\%. The
uncertainties in the b-tagging efficiency and mistagging probability
depend on the $\eta$ and \pt of the jet and are on average 4\% and 10\%,
respectively \cite{CMS-PAS-BTV-13-001}.

Systematic uncertainties, totaling 19--22\% depending on the channel and the search, are assigned to
the normalization of the \ttbar irreducible background based on statistical
uncertainty in the control samples and the propagation of the uncertainties
in the acceptances, efficiencies, and subtraction of the contributions from other processes in the \emu sample.
Systematic uncertainties in the major reducible background are driven by statistical uncertainty
in the measured misidentification probability and variation in the misidentification probability
based on the event topology. These uncertainties amount to 16--24\%,
depending on the channel and the search.

Because of the limited number of events in the simulation, uncertainties
in the small backgrounds range between 20--50\%.
Uncertainty due to the effect of pileup modeling in the MC is
estimated to be 3\%.
A 4\% uncertainty, due to modeling of initial- and final-state radiation
in the simulation, is assigned to the signal acceptance.
The uncertainty in the initial- and final-state radiation 
was found to have a negligible effect on the simulated backgrounds.
A 7--32\% uncertainty from knowledge of parton distribution functions and a 14--80\%
uncertainty from QCD renormalization and factorization scales are assigned to the
theoretical signal cross-section.
Finally, jet energy scale uncertainties (2--4\%
depending on $\eta$ and \pt)
and energy resolution uncertainties (5--10\% depending on $\eta$),
as well as energy scale (3\%) and resolution (10\%) uncertainties
for $\tauh$, affect both the \ST distributions
and the expected yields
from the signal and background processes.

\section{Results}

\begin{figure}[htbp]
  \centering
    \includegraphics[width=\cmsFigWidth]{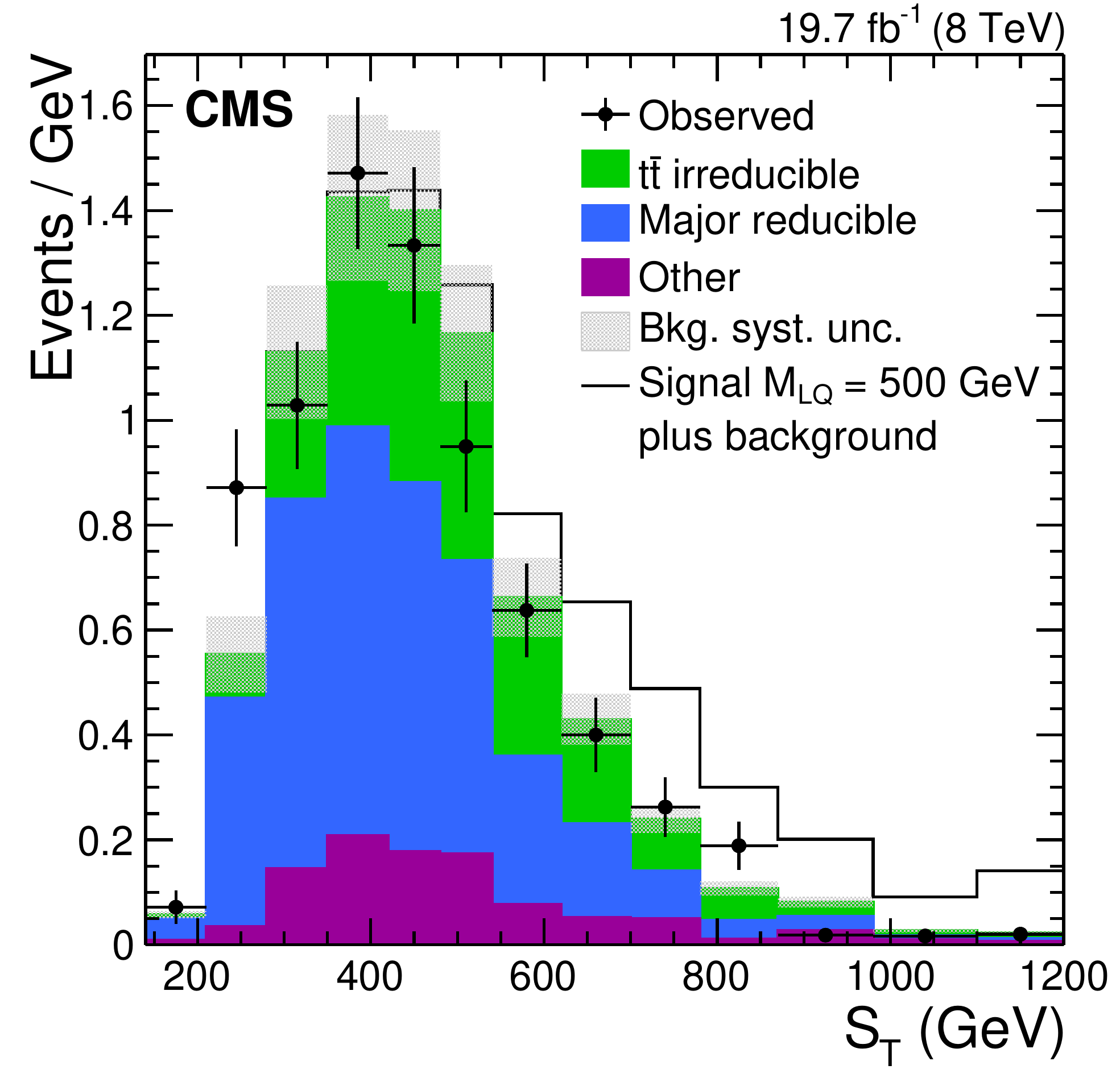}
    \caption{The final \ST distribution for the leptoquark search with the \etau and \mutau channels combined.
             A signal sample for leptoquarks with the mass of 500\GeV is added on top of the background prediction.
             The last bin contains the overflow events. The horizontal bar on each observed data point indicates the width of the bin in \ST.
           }
    \label{Res:fig:STfinalLQ}
\end{figure}

\begin{figure}[htbp]
  \centering
    \includegraphics[width=\cmsFigWidth]{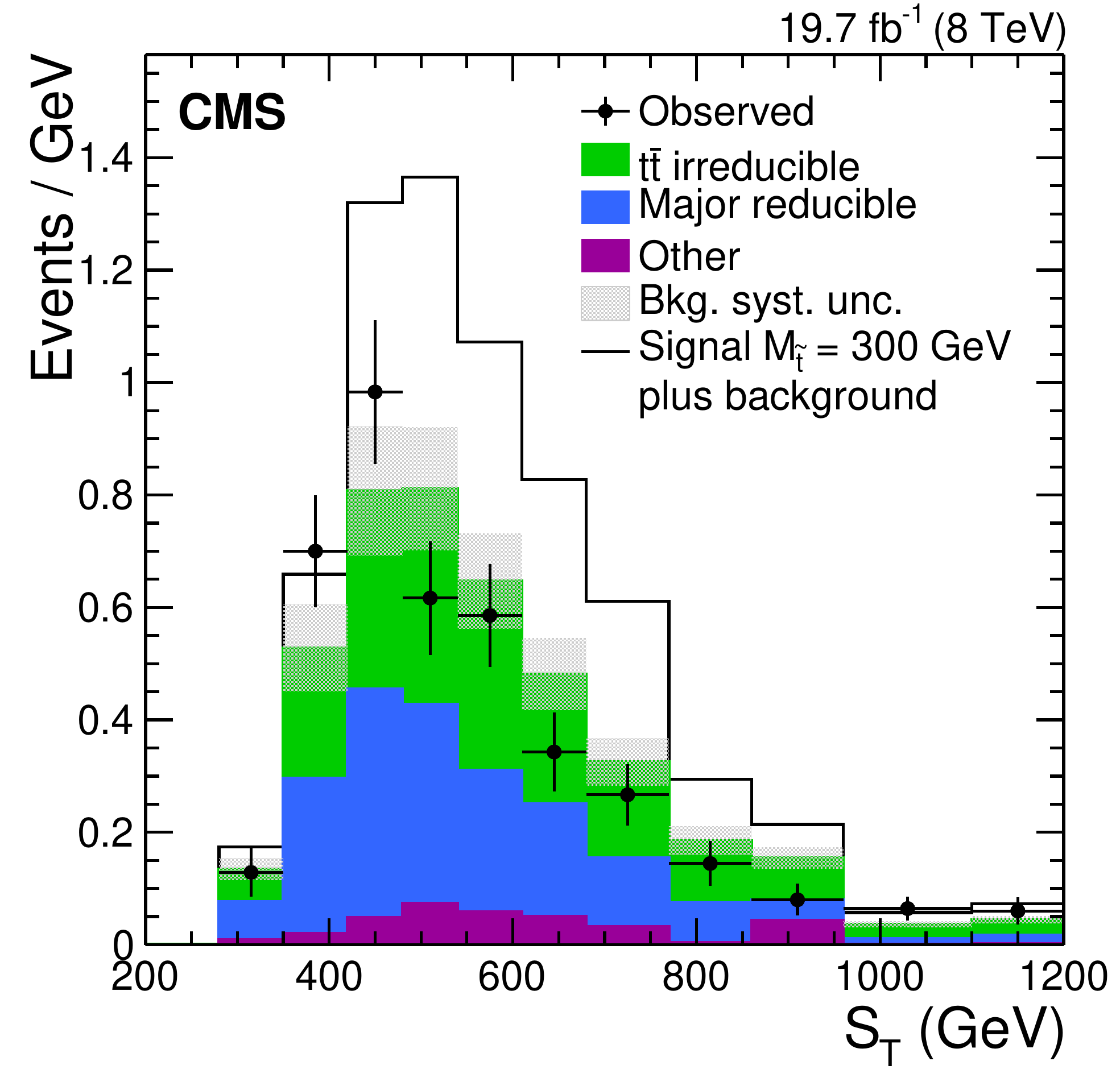}
    \caption{The final \ST distribution for the top squark search with the \etau and \mutau channels combined.
             A signal sample for top squarks with the mass of 300\GeV is added on top of the background prediction.
             The last bin contains the overflow events. The horizontal bar on each observed data point indicates the width of the bin in \ST.
           }
    \label{Res:fig:STfinalLQD321}
\end{figure}

The numbers of observed events and expected signal and background events after the final
selection for the leptoquark and top squark searches are listed in Tables~\ref{Res:tab:STyieldLQ}
and \ref{Res:tab:STyieldLQD321}, respectively, and the selection efficiencies for the two signals are listed in
Tables~\ref{Res:tab:effLQ} and \ref{Res:tab:effLQD321}.
The \ST distributions of the selected events from the observed data and from the background predictions, combining
\etau and \mutau channels, are shown in Fig. \ref{Res:fig:STfinalLQ}
for the leptoquark search
and Fig. \ref{Res:fig:STfinalLQD321} for the top squark search.
The distribution from the 500\GeV (300\GeV) signal hypothesis is added to the background
in Fig. \ref{Res:fig:STfinalLQ} (Fig. \ref{Res:fig:STfinalLQD321})
to illustrate how a hypothetical signal would appear above the background prediction.
The data agree well with the SM background prediction.

\begin{figure}[htbp]
  \centering
    \includegraphics[width=\cmsFigWidth]{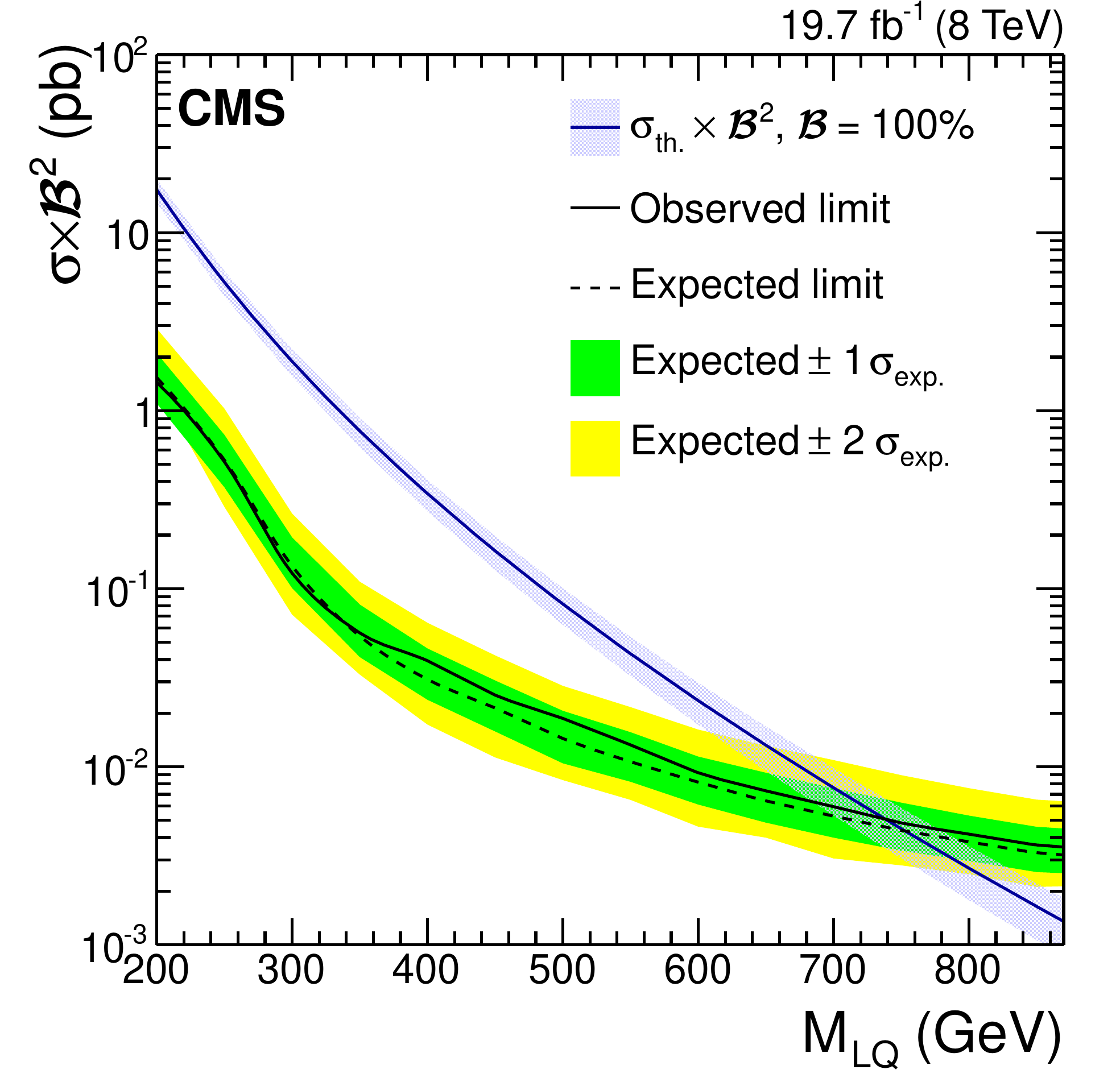}
    \caption{The expected and observed combined upper limits on the third-generation LQ pair production cross section $\sigma$ times the square of the branching fraction, $\mathcal{B}^2$, at 95\% CL, as a function of the LQ mass. These limits also apply to top squarks decaying directly via the coupling $\lambda^{\prime}_{333}$. The green (darker) and yellow (lighter) uncertainty bands represent 68\% and 95\% CL intervals on the expected limit. The dark blue curve and the hatched light blue band represent the theoretical LQ pair production cross section, assuming $\mathcal{B}=100\%$, and the uncertainties due to the choice of PDF and renormalization/factorization scales.}
    \label{Res:fig:asymptoticCombLQ}
\end{figure}

\begin{figure}[htbp]
\centering
    \includegraphics[width=\cmsFigWidth]{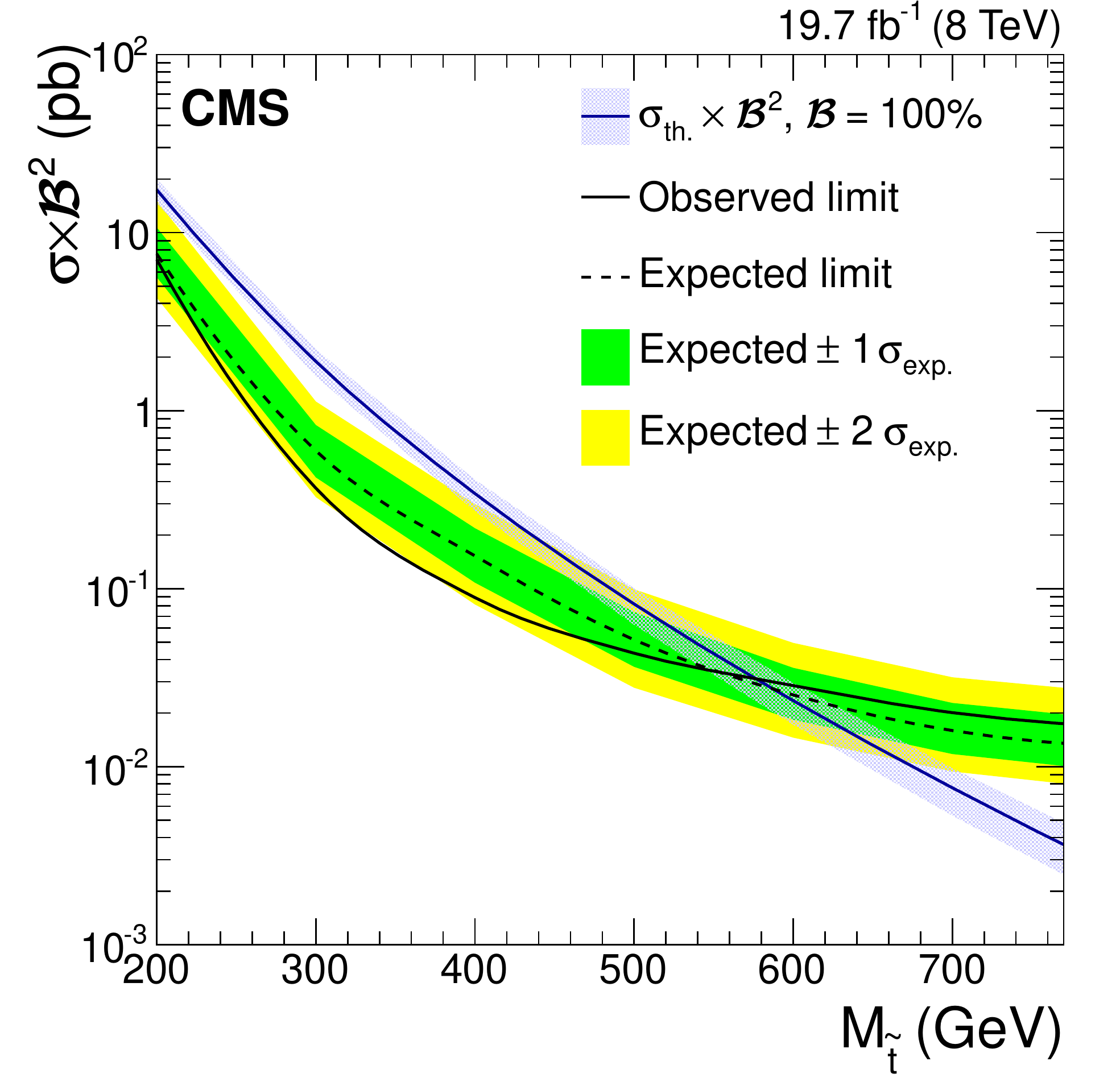}
    \caption{The expected and observed combined upper limits on the top squark pair production cross section $\sigma$ times the square of the branching fraction, $\mathcal{B}^2$, at 95\% CL, as a function of the top squark mass. These limits apply to top squarks with a chargino-mediated decay through the coupling $\lambda^{\prime}_{3kj}$. The green (darker) and yellow (lighter) uncertainty bands represent 68\% and 95\% CL intervals on the expected limit. The dark blue curve and the hatched light blue band represent the theoretical top squark pair production cross section, assuming $\mathcal{B}=100\%$, and the uncertainties due to the choice of PDF and renormalization/factorization scales.}
    \label{Res:fig:asymptoticCombLQD}
\end{figure}

An upper bound at 95\% confidence level (CL) is set on $\sigma \mathcal{B}^2$, where
$\sigma$ is the cross section for pair production of
third-generation LQs (top squarks)
and $\mathcal{B}$ is the branching fraction
for the LQ decay to a tau lepton and a bottom quark
(the top squark decay to a $\chipm$ and a bottom quark, with a
subsequent decay of the chargino via
$\chipm\to\sNu\tau^{\pm} \to \cPq\cPq\tau^{\pm}$).
The symbol $M_{\text{LQ}}$ is used for the leptoquark mass and
the symbol $M_{\sTop}$ is used for the top squark mass.
The modified-frequentist construction
CL$_\mathrm{s}$~\cite{Read,Junk,LHC-HCG}
is used for the limit calculation.
A maximum likelihood fit is performed to
the \ST spectrum simultaneously for
the \etau and \mutau channels,
taking into account
correlations between the systematic uncertainties.
Expected  and observed upper limits on $\sigma \mathcal{B}^2$
as a function of the signal mass are shown in
Fig.~\ref{Res:fig:asymptoticCombLQ} for the leptoquark search and
Fig.~\ref{Res:fig:asymptoticCombLQD} for the top squark search.

We extend the current limits and exclude scalar leptoquarks and top squarks decaying through the coupling $\lambda^{\prime}_{333}$
with masses below 740\GeV, in agreement with a limit at 750\GeV, expected in the absence of a signal.
We exclude top squarks undergoing a chargino-mediated decay involving the coupling $\lambda^{\prime}_{3jk}$
with masses in the range 200--580\GeV, in agreement
with the expected exclusion limit in the range 200--590\GeV.
These upper limits assume $\mathcal{B}=100\%$.
Similar results are obtained when calculating upper bounds using a
Bayesian method with a uniform positive prior for the cross section.

\begin{figure}[htbp]
\centering
    \includegraphics[width=\cmsFigWidth]{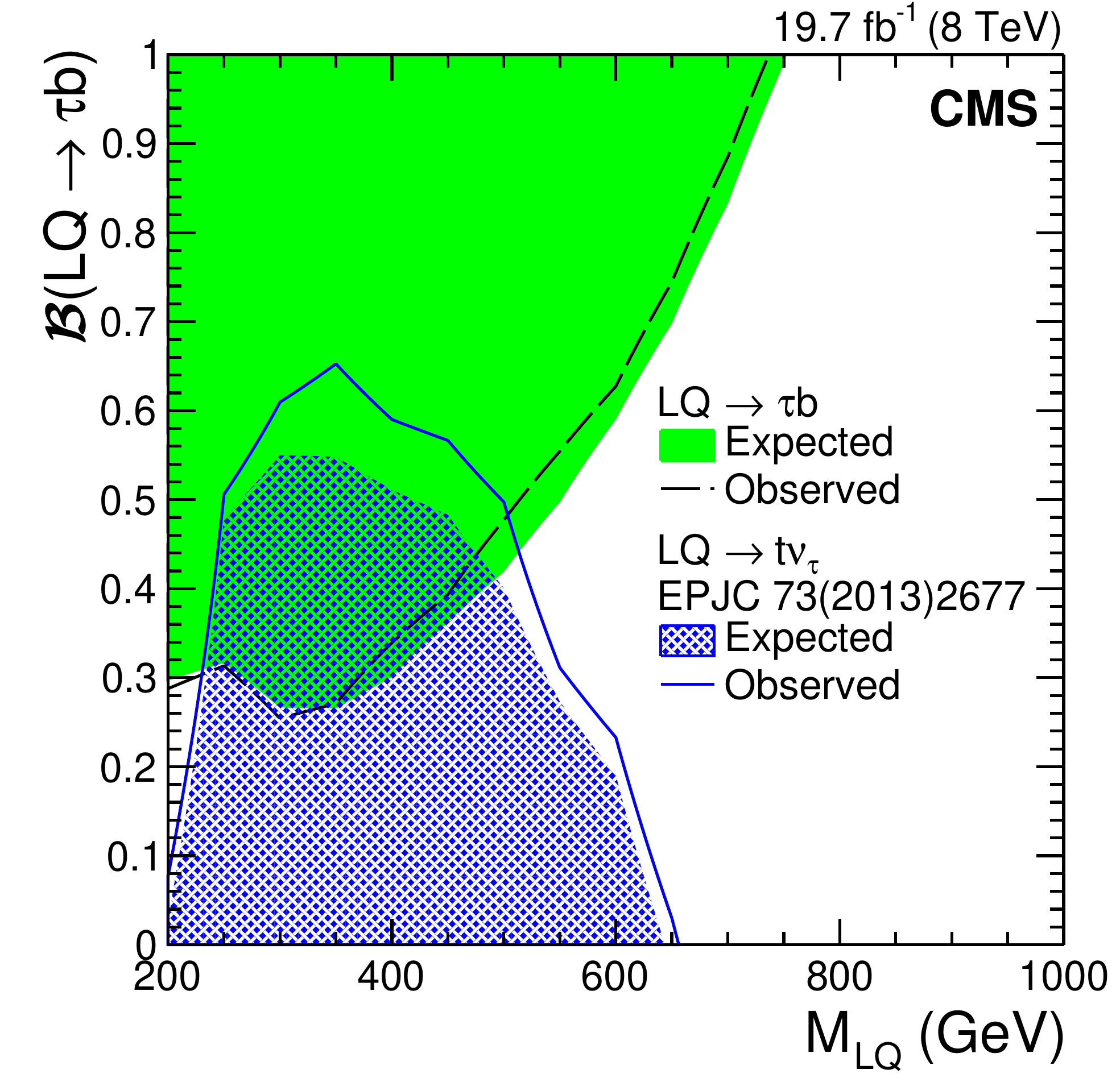}
    \caption{The expected (dashed black) and observed (green solid) 95\% CL upper limits on the branching fraction for the leptoquark decay to a tau lepton and a bottom quark, as a function of the leptoquark mass. A search for top squark pair production \cite{SUS-13-011} has the same kinematic signature as the leptoquark decay to a tau neutrino and a top quark. This search is reinterpreted to provide the expected (blue hatched) and observed (blue open) 95\% CL upper limits for low values of $\mathcal{B}$, assuming the leptoquark only decays to third-generation fermions.}
    \label{Res:fig:2DCombLQ}
\end{figure}

The upper bounds for the leptoquark search as a function of the leptoquark branching fraction
and mass are shown in Fig.~\ref{Res:fig:2DCombLQ}.
Small $\mathcal{B}$ values are not constrained by this search.
Results from the CMS experiment on a search for top squarks
decaying to a top quark and a neutralino~\cite{SUS-13-011} are used to further constrain $\mathcal{B}$.
If the neutralino is massless, the final state kinematic distributions for such a signal are the same
as those for the pair production of leptoquarks decaying to a tau neutrino and a
top quark. Limits can therefore be placed on this signal,
which must have a branching fraction of $1-\mathcal{B}$
if the leptoquark only decays to third-generation fermions.
This reinterpretation is included in Fig.~\ref{Res:fig:2DCombLQ}.
The unexcluded region at $M_{\text{LQ}}=200$--230\GeV
corresponds to a portion of phase space where it
is topologically very difficult to distinguish
between the top squark signal
and the \ttbar process, owing to small missing transverse momentum.
A top squark excess in this region would imply an excess in the measured
\ttbar cross section of $\sim$10\%.

\begin{table}[htbp]
  \centering
    \topcaption{The estimated backgrounds, observed event yields, and expected number of signal events for the leptoquark search. For the simulation-based entries, the statistical and systematic uncertainties are shown separately, in that order.}
    \begin{tabular}{lcc}
      \hline
      & $\etau$ & $\mutau$ \\
      \cline{2-3}
      \ttbar irreducible              &  105.6$\pm$18.1           &   66.7$\pm$12.6  	    \\
      Major reducible                 &  147.8$\pm$33.0           &  117.3$\pm$18.9  	    \\
      Z($\ell\ell$/$\tau\tau$)+jets   &    21.4$\pm$7.4$\pm$4.9   &    7.5$\pm$4.6$\pm$0.2  \\
      Single t                        &   16.0$\pm$2.8$\pm$4.4    &   17.3$\pm$2.8$\pm$4.7  \\
      VV                              &   4.1$\pm$0.6$\pm$1.3     &    2.6$\pm$0.5$\pm$0.8  \\
      \hline
      Total exp. bkg.                 & 294.9$\pm$7.9$\pm$39.1 & 211.4$\pm$5.4$\pm$23.4   \\
      \hline
      Observed                       & 289                         & 216                      \\
      \hline
     $M_{\text{LQ}}=500\GeV$            & 57.7$\pm$1.4$\pm$5.9      & 51.6$\pm$1.3$\pm$5.3    \\
     $M_{\text{LQ}}=600\GeV$            & 20.1$\pm$0.5$\pm$1.9      & 17.7$\pm$0.4$\pm$1.6    \\
     $M_{\text{LQ}}=700\GeV$            & 7.1$\pm$0.2$\pm$6.3       & 6.2$\pm$0.1$\pm$5.5    \\
     $M_{\text{LQ}}=800\GeV$            & 2.7$\pm$0.1$\pm$0.2       & 2.3$\pm$0.1$\pm$0.2    \\
      \hline
    \end{tabular}
    \label{Res:tab:STyieldLQ}
\end{table}

\begin{table}[htbp]
  \centering
    \topcaption{The estimated backgrounds, observed event yields, and expected number of signal events for the top squark search. For the simulation-based entries, the statistical and systematic uncertainties are shown separately, in that order.}
    \begin{tabular}{lcc}
      \hline
      & $\etau$ & $\mutau$ \\
      \cline{2-3}
      \ttbar irreducible             & 88.3$\pm$13.7            &  55.0$\pm$9.5           \\
      Major reducible                & 65.7$\pm$16.4            &  59.8$\pm$13.8          \\
      Z($\ell\ell$/$\tau\tau$)+jets  & 4.9$\pm$2.5$\pm$1.1      &  11.6$\pm$5.5$\pm$2.7   \\
      Single t                       & 3.9$\pm$1.5$\pm$1.1      &  3.5$\pm$1.3$\pm$0.9   \\
      VV                             & 0.6$\pm$0.2$\pm$0.2       &  0.4$\pm$0.2$\pm$0.1   \\
      \hline
      Total exp. bkg.            & 163.4$\pm$2.9$\pm$21.5  & 130.3$\pm$5.6$\pm$17.1 \\
      \hline
      Observed                     & 156                        & 123                    \\
      \hline
      $M_{\sTop}=300\GeV$           & 94.3$\pm$8.5$\pm$13.2  &  82.8$\pm$8.0$\pm$11.7  \\
      $M_{\sTop}=400\GeV$           & 43.9$\pm$2.6$\pm$4.3   &  38.3$\pm$2.3$\pm$3.8   \\
      $M_{\sTop}=500\GeV$           & 19.4$\pm$0.8$\pm$1.8   &  15.4$\pm$0.7$\pm$1.5   \\
      $M_{\sTop}=600\GeV$           & 6.9$\pm$0.9$\pm$0.7    &   5.7$\pm$0.3$\pm$0.5   \\
      \hline
    \end{tabular}
    \label{Res:tab:STyieldLQD321}
\end{table}

\begin{table}[htbp]
  \centering
    \topcaption{Selection efficiencies in \% for the signal in the leptoquark search, estimated from the simulation.}
    \begin{tabular}{lll}
\hline
$M_{\text{LQ}}$ (\GeVns) & \etau & \mutau \\
\hline
200  & 0.1 & 0.1 \\
250  & 0.3 & 0.2 \\
300  & 1.0 & 0.8 \\
350  & 1.9 & 1.5 \\
400  & 2.4 & 2.3 \\
450  & 3.0 & 2.9 \\
500  & 3.6 & 3.2 \\
550  & 4.0 & 3.3 \\
600  & 4.4 & 3.8 \\
650  & 4.5 & 4.0 \\
700  & 4.7 & 4.1 \\
750  & 4.9 & 4.2 \\
800  & 5.1 & 4.3 \\
850  & 5.4 & 4.4 \\
900  & 5.1 & 4.4 \\
950  & 5.4 & 4.3 \\
1000 & 5.5 & 4.4 \\
\hline
    \end{tabular}
    \label{Res:tab:effLQ}
\end{table}

\begin{table}[htbp]
  \centering
    \topcaption{Selection efficiencies in \% for the signal in the top squark search, estimated from the simulation.}
    \begin{tabular}{lll}
\hline
$M_{\sTop}$ (\GeVns) & \etau & \mutau \\
\hline
200 & 0.02 & 0.02 \\
300 & 0.3 & 0.2 \\
400 & 0.7 & 0.6 \\
500 & 1.2 & 1.0 \\
600 & 1.5 & 1.2 \\
700 & 1.8 & 1.4 \\
800 & 1.8 & 1.3 \\
900 & 1.5 & 1.1 \\
\hline
    \end{tabular}
    \label{Res:tab:effLQD321}
\end{table}

\section{Summary}

A search for pair production of third-generation
scalar leptoquarks and top squarks has been presented.
The search for leptoquarks and top squarks decaying
through the R-parity violating coupling $\lambda^{\prime}_{333}$
is performed in final states that include
an electron or a muon, a hadronically decaying tau lepton,
and at least two jets, at least one of which is b-tagged.
The search for top squarks undergoing a chargino-mediated decay
involving the R-parity violating coupling $\lambda^{\prime}_{3jk}$
is performed in events containing
an electron or a muon, a hadronically decaying tau lepton,
and at least five jets, at least one of which is b-tagged.
No excesses above the standard model background prediction are observed in the \ST distributions.
Assuming a 100\% branching fraction for the
decay to a tau lepton and a bottom quark,
scalar leptoquarks and top squarks decaying through $\lambda^{\prime}_{333}$ with masses below 740\GeV
are excluded at 95\% confidence level. Top squarks decaying through $\lambda^{\prime}_{3jk}$ with masses below 580\GeV
are excluded at 95\% confidence level, assuming a 100\% branching fraction for the decay to a tau lepton, a bottom quark,
and two light quarks.
The constraint on the third-generation leptoquark mass is the most stringent to date,
and this is the first search for top squarks decaying through $\lambda^{\prime}_{3jk}$.

\section*{Acknowledgements}

We congratulate our colleagues in the CERN accelerator departments for the excellent performance of the LHC and thank the technical and administrative staffs at CERN and at other CMS institutes for their contributions to the success of the CMS effort. In addition, we gratefully acknowledge the computing centres and personnel of the Worldwide LHC Computing Grid for delivering so effectively the computing infrastructure essential to our analyses. Finally, we acknowledge the enduring support for the construction and operation of the LHC and the CMS detector provided by the following funding agencies: BMWFW and FWF (Austria); FNRS and FWO (Belgium); CNPq, CAPES, FAPERJ, and FAPESP (Brazil); MES (Bulgaria); CERN; CAS, MoST, and NSFC (China); COLCIENCIAS (Colombia); MSES and CSF (Croatia); RPF (Cyprus); MoER, ERC IUT and ERDF (Estonia); Academy of Finland, MEC, and HIP (Finland); CEA and CNRS/IN2P3 (France); BMBF, DFG, and HGF (Germany); GSRT (Greece); OTKA and NIH (Hungary); DAE and DST (India); IPM (Iran); SFI (Ireland); INFN (Italy); NRF and WCU (Republic of Korea); LAS (Lithuania); MOE and UM (Malaysia); CINVESTAV, CONACYT, SEP, and UASLP-FAI (Mexico); MBIE (New Zealand); PAEC (Pakistan); MSHE and NSC (Poland); FCT (Portugal); JINR (Dubna); MON, RosAtom, RAS and RFBR (Russia); MESTD (Serbia); SEIDI and CPAN (Spain); Swiss Funding Agencies (Switzerland); MST (Taipei); ThEPCenter, IPST, STAR and NSTDA (Thailand); TUBITAK and TAEK (Turkey); NASU and SFFR (Ukraine); STFC (United Kingdom); DOE and NSF (USA).

Individuals have received support from the Marie-Curie programme and the European Research Council and EPLANET (European Union); the Leventis Foundation; the A. P. Sloan Foundation; the Alexander von Humboldt Foundation; the Belgian Federal Science Policy Office; the Fonds pour la Formation \`a la Recherche dans l'Industrie et dans l'Agriculture (FRIA-Belgium); the Agentschap voor Innovatie door Wetenschap en Technologie (IWT-Belgium); the Ministry of Education, Youth and Sports (MEYS) of the Czech Republic; the Council of Science and Industrial Research, India; the HOMING PLUS programme of Foundation for Polish Science, cofinanced from European Union, Regional Development Fund; the Compagnia di San Paolo (Torino); the Consorzio per la Fisica (Trieste); MIUR project 20108T4XTM (Italy); the Thalis and Aristeia programmes cofinanced by EU-ESF and the Greek NSRF; and the National Priorities Research Program by Qatar National Research Fund.

\bibliography{auto_generated}
\cleardoublepage \appendix\section{The CMS Collaboration \label{app:collab}}\begin{sloppypar}\hyphenpenalty=5000\widowpenalty=500\clubpenalty=5000\textbf{Yerevan Physics Institute,  Yerevan,  Armenia}\\*[0pt]
V.~Khachatryan, A.M.~Sirunyan, A.~Tumasyan
\vskip\cmsinstskip
\textbf{Institut f\"{u}r Hochenergiephysik der OeAW,  Wien,  Austria}\\*[0pt]
W.~Adam, T.~Bergauer, M.~Dragicevic, J.~Er\"{o}, C.~Fabjan\cmsAuthorMark{1}, M.~Friedl, R.~Fr\"{u}hwirth\cmsAuthorMark{1}, V.M.~Ghete, C.~Hartl, N.~H\"{o}rmann, J.~Hrubec, M.~Jeitler\cmsAuthorMark{1}, W.~Kiesenhofer, V.~Kn\"{u}nz, M.~Krammer\cmsAuthorMark{1}, I.~Kr\"{a}tschmer, D.~Liko, I.~Mikulec, D.~Rabady\cmsAuthorMark{2}, B.~Rahbaran, H.~Rohringer, R.~Sch\"{o}fbeck, J.~Strauss, A.~Taurok, W.~Treberer-Treberspurg, W.~Waltenberger, C.-E.~Wulz\cmsAuthorMark{1}
% ...
\vskip\cmsinstskip
% ...
1:~~Also at Vienna University of Technology, Vienna, Austria\\
2:~~Also at CERN, European Organization for Nuclear Research, Geneva, Switzerland\\
% ...

\end{sloppypar}
\end{document}